\documentclass[fleqn,usenatbib]{mnras}

\usepackage{newtxtext,newtxmath}

\usepackage[T1]{fontenc}
\usepackage{ae,aecompl}

\usepackage{graphicx}	
\usepackage{amsmath}	
\usepackage{amssymb}	
\usepackage{caption,subcaption,tikz}
\usepackage{soul}
\sethlcolor{yellow}
\soulregister\citet7
\soulregister\citep7
\soulregister\citealp7
\soulregister\ref7
\soulregister\url7
\usepackage{breakcites}

\title[Revisiting hypervelocity stars]{Revisiting hypervelocity stars after Gaia DR2}

\author[D. Boubert et al.]{
D. Boubert,$^{1}$\thanks{E-mail: d.boubert@ast.cam.ac.uk}
J. Guillochon,$^{2}$\thanks{E-mail: jguillochon@cfa.harvard.edu}
K. Hawkins,$^{3}$
I. Ginsburg,$^{2}$
N. W. Evans,$^{1}$ and
\newauthor
J. Strader$^{4}$
\\
$^{1}$Institute of Astronomy, University of Cambridge, Madingley Road, Cambridge CB3 0HA, UK\\
$^{2}$Harvard-Smithsonian Center for Astrophysics, 60 Garden St., Cambridge, MA 02138, USA\\
$^{3}$Department of Astronomy, the University of Texas at Austin, 2515 Speedway Boulevard, Austin, TX 78712, USA\\
$^{4}$Department of Physics and Astronomy, Michigan State University, East Lansing, MI 48824, USA
}

\date{Accepted XXX. Received YYY; in original form ZZZ}

\pubyear{2018}

\begin{document}
\label{firstpage}
\pagerange{\pageref{firstpage}--\pageref{lastpage}}
\maketitle

\begin{abstract}
Hypervelocity stars are intriguing rare objects traveling at speeds large enough to be unbound from the Milky Way. Several mechanisms have been proposed for producing them, including the interaction of the Galaxy's super-massive black hole (SMBH) with a binary; rapid mass-loss from a companion to a star in a short-period binary; the tidal disruption of an infalling galaxy and finally ejection from the Large Magellanic Cloud. While previously discovered high-velocity early-type stars are thought to be the result of an interaction with the SMBH, the origin of high-velocity late type stars is ambiguous. The second data release of Gaia (DR2) enables a unique opportunity to resolve this ambiguity and determine whether any late-type candidates are truly unbound from the Milky Way. In this paper, we utilize the new proper motion and velocity information available from DR2 to re-evaluate a collection of historical data compiled on the newly-created Open Fast Stars Catalog. We find that almost all previously-known high-velocity late-type stars are most likely bound to the Milky Way. Only one late-type object (LAMOST J115209.12+120258.0) is unbound from the Galaxy. Performing integrations of orbital histories, we find that this object cannot have been ejected from the Galactic centre and thus may be either debris from the disruption of a satellite galaxy or a disc runaway.
\end{abstract}




\section{Introduction}
Gaia has ushered in the era of Billion Star Maps of the Milky Way's dynamics, yet we do not know the origin or history for the fastest stars in the Galaxy. These unbound stars are defined as having a speed above the escape speed of the Galaxy at their location and are commonly referred to as `hypervelocity stars' \citep[e.g.][]{hills_hyper-velocity_1988,brown_hypervelocity_2015}. The first hypervelocity star (later denoted HVS1) was serendipitously discovered by \citet{brown_discovery_2005} and found to be a $3\;\mathrm{M}_{\odot}$ late B-type star with a heliocentric distance $71\;\mathrm{kpc}$ and radial velocity $853\pm12\;\mathrm{km}\;\mathrm{s}^{-1}$. The hypervelocity classification of HVS1 is secure, because i) the radial velocity is sufficient to make the star unbound even without adding on the proper motion and ii) a B-type star could only reach the outer halo if it had such an extreme velocity.

\begin{figure*}
\includegraphics[width=0.8\linewidth]{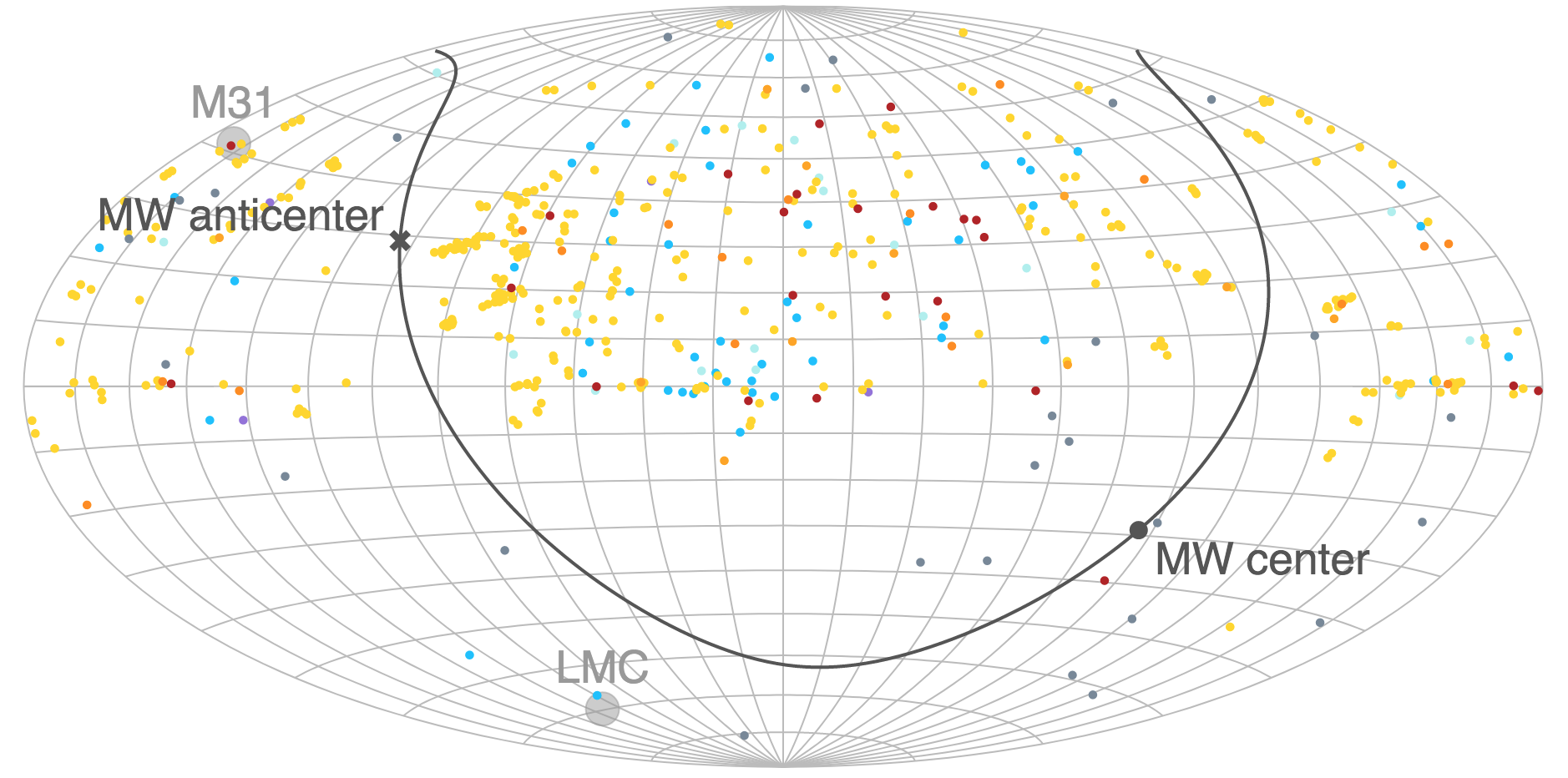}\\
Spectral class:\\
\includegraphics[width=0.55\linewidth]{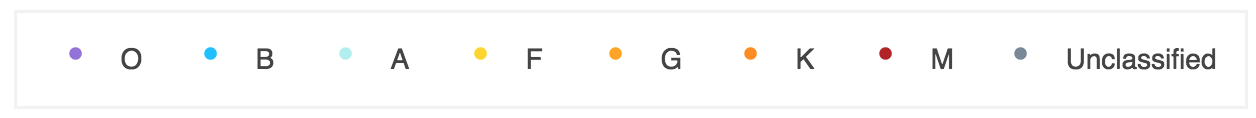}
    \caption{Hammer projection in right ascension and declination of all previously-known high-velocity stars, color-coded by spectral type. The thick gray line shows the plane of the Milky Way, with the large gray dot indicating the location of the galactic center. The locations of M31 and the LMC are shown as annotated. An interactive version of this figure is available at the OFSC\protect\footnotemark.}
    \label{fig:skymap}
\end{figure*}

The number of candidate hypervelocity stars has ballooned in the years since the discovery of HVS1 and today there are more than 500 candidates in the literature\footnote{\url{https://faststars.space}} (see Fig. \ref{fig:skymap}). There is, however, reason to be skeptical of many of these candidates. While \citet{brown_hypervelocity_2006,zheng_first_2014,brown_mmt_2014,huang_discovery_2017} have discovered a further two dozen hypervelocity candidates that are likely late B-type stars far out in the halo and with an extreme radial velocity, most of the candidates are late-type, high proper motion stars. In a majority of cases, the radial velocity is itself unremarkable and the `hypervelocity' classification is driven entirely by a large proper motion measurement. However, as noted in \citet{ziegerer_candidate_2015}, there is reason to be cautious. The authors assessed the 
candidates in \citet{palladino_hypervelocity_2014}. They were unable to confirm them, with the ground-based proper motions fingered as the likely culprit. 

The origin of hypervelocity stars remains an intriguing and open question. The tidal disruption of binary stars by the supermassive black hole at the Galactic centre, leading to the ejection of one of the stars \citep{hills_hyper-velocity_1988}, is considered the most likely possibility \citep[e.g.][]{ginsburg_hypervelocity_2007,brown_hypervelocity_2015} . However, there remains the possibility that the hypervelocity stars were ejected from elsewhere in the Milky Way's disc and are either supernova runaways or were dynamically ejected from star clusters. Recently \citet{boubert_dipole_2016} and \citet{boubert_hypervelocity_2017} argued that the hypervelocity stars could possibly originate in the Large Magellanic Cloud. The early-type hypervelocity stars are found in the halo, while the late-type hypervelocity stars are found within several kiloparsecs of the Sun. Thus, these two populations probe different kinematic regimes and can potentially be used to distinguish between the formation scenarios. The question of whether there are any late-type hypervelocity stars lies at the centre of the hypervelocity star mystery.

The European Space Agency's Gaia space telescope was launched in 2013 and on the 25$^{\mathrm{th}}$ April 2018 delivered its second date release \citep[Gaia DR2,][]{gaia_collaboration_gaia_2016,2018arXiv180409365G}  containing astrometry and photometry for 1,692,919,135 sources, based on the first 22 months of operation. This catalogue includes parallaxes and proper motions for an unprecendented 1,331,909,727 sources, typically with sub-milliarcsecond precision\footnote{\url{https://www.cosmos.esa.int/web/gaia/data}}. Gaia can thus revolutionise the study of late-type hypervelocity stars. It will allow accurate tangential velocities to be obtained for all extant late-type hypervelocity candidates.

The objective of this paper is to provide a comprehensive update on the status of the hypervelocity candidates in the literature after Gaia DR2. We specifically focus on the nearby, late-type candidates because these are the stars whose status is most likely to change with improved astrometry. In Section \ref{sec:history}, we briefly cover the history of searches for late-type hypervelocity stars. Section \ref{sec:results} provides an overview of the landscape of hypervelocity star candidates and looks in detail at the one confirmed late-type hypervelocity star. In the Conclusions, we discuss the implications of our results. In the Appendix we present the Open Fast Stars Catalog whose creation enabled this work.

\section{History of searches for late-type hypervelocity stars}
\label{sec:history}

Prior to Gaia DR2, a number of late-type hypervelocity candidates had been claimed in the literature. We define late-type as stars whose spectral type is F, G, K or M, including both dwarf and giant stars. Many of these identifications were based on cross-matches between spectroscopic surveys such as SEGUE~\citep{SEGUE} and LAMOST~\citep{LAMOST} together with the SDSS-USNO proper motion catalogues~\citep{munn_improved_2004,Mu08}. With the addition of photometric parallaxes, this gives the full space motion of the candidate. The orbit is integrated in a Galactic model to assess whether it is unbound. The radial velocity is usually secure, but photometric parallaxes typically have errors of $\sim 15$ per cent. Even the most carefully constructed ground-based proper motion catalogues tend to have some erroneous measurements, especially in the high proper motion regime.

As an example, \citet{Li12} searched through Sloan Digital Sky Survey (SDSS) Data Release 7 and identified 13 F-type hypervelocity star candidates. They used SEGUE spectroscopy and proper motions from the SDSS-USNO~\citep{munn_improved_2004}. They argued from orbit integrations that 9 candidates emanated from the Galactic Center of disk, whilst the remaining 4 had a more exotic origin, such as tidal disruption of dwarf galaxies~\citep{Ab09}. \citet{palladino_hypervelocity_2014} also carried out a search in the SEGUE G and K dwarfs sample, again based on proper motions from SDSS+USNO-B \citep{munn_improved_2004}. The fate of these candidates illustrates the pitfalls of such work. Many of the candidates were contested either because they are high velocity halo stars and therefore bound or because the ground-based proper motions are inflated~\citep{ziegerer_candidate_2015}.

The LAMOST survey also proved to be a happy hunting ground for late-type hypervelocity star candidates. \citet{li_19_2015} claimed 19 low mass F, G and K type hypervelocity star candidates from over one million stars found in the first data release of the LAMOST regular survey. They combined LAMOST spectroscopy with SDSS-USNO-B~\citep{Mu08} proper motions. Their final cleaned candidate list used only stars with reliable proper motions, high quality spectra and trustworthy astrophysical parameters. The candidates had probabilities of being unbound, as judged from Monte Carlo simulations of orbit integrations, in excess of 50 per cent. However, there were 8 high quality candidates with a probability in excess of 80 per cent.

We are not the first to realise the potential of Gaia as a purger of late-type hypervelocity candidates.  \citet{marchetti_artificial_2017} trained a neural network to identify hypervelocity star candidates in Gaia DR1 and noticed that one of their candidates HD 5223 had previously been suggested by \citet{pereira_cd-621346:_2012}. The Gaia parallax indicated that it was much closer than previously thought. Given the history of the subject, Gaia Data Release 2 proper motions might well be expected to winnow the late-type hypervelocity candidates.

\begin{figure*}
	\begin{subfigure}{0.8\textwidth}
		\includegraphics[width=\linewidth]{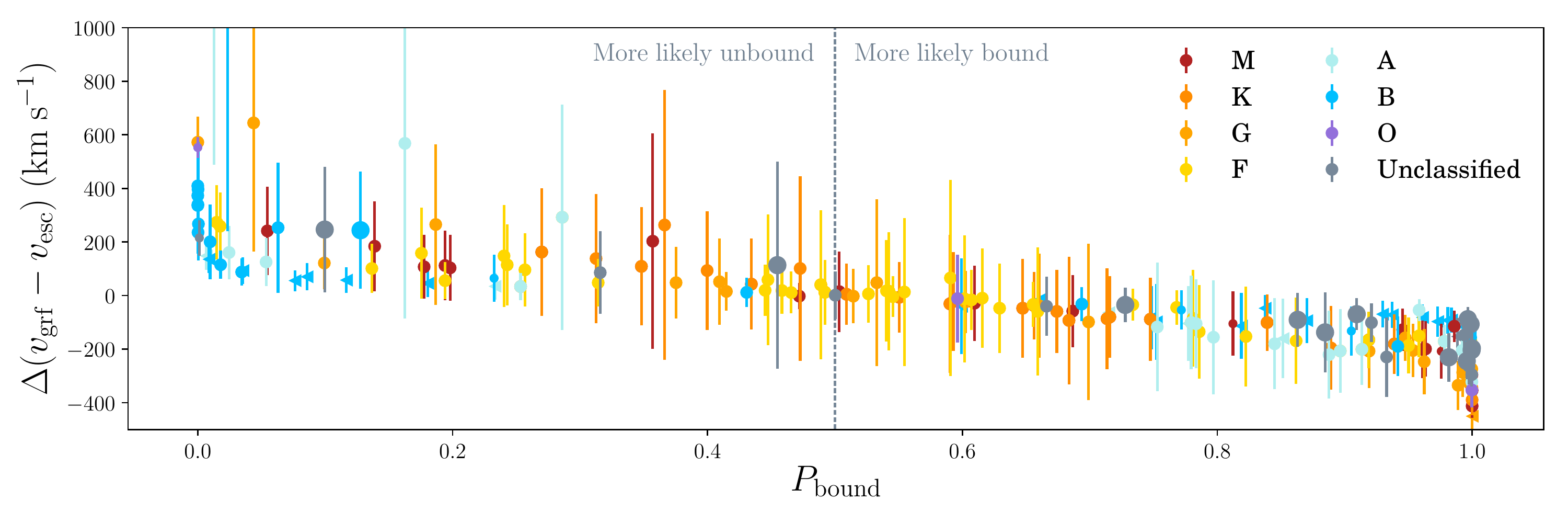}
		\caption{$P_{\mathrm{bound}}$ before Gaia DR2}
		\label{fig:pboundbefore}
	\end{subfigure}
	\begin{subfigure}{0.8\textwidth}
		\includegraphics[width=\linewidth]{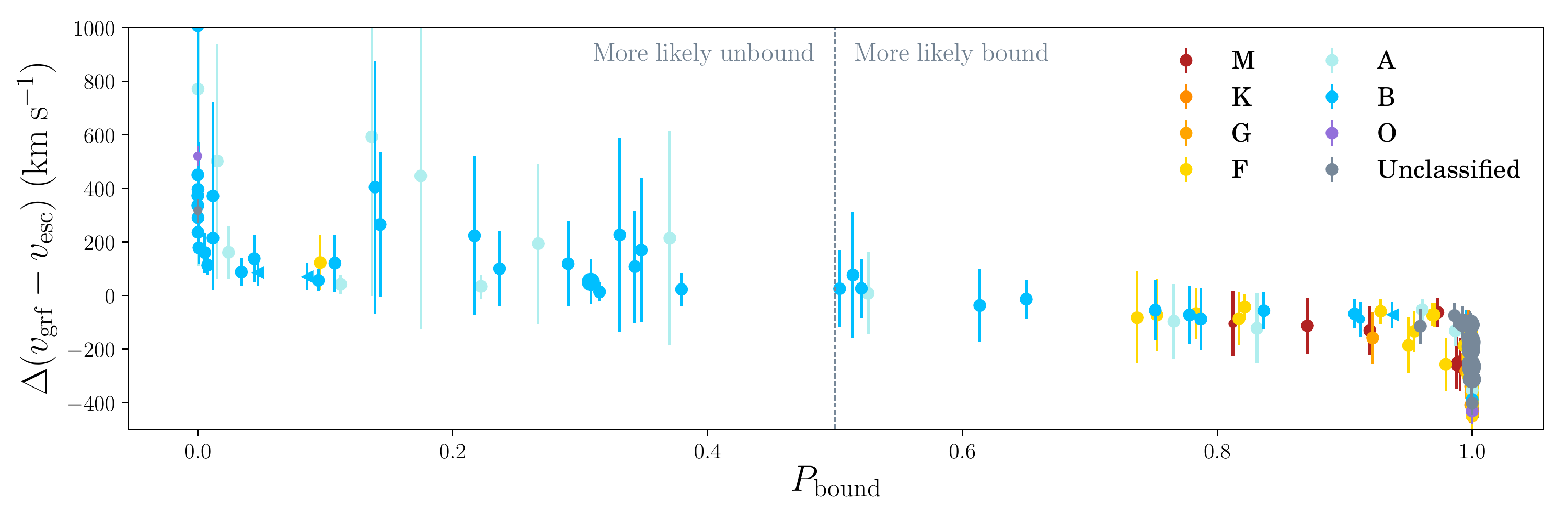}
		\caption{$P_{\mathrm{bound}}$ after Gaia DR2}
		\label{fig:pboundafter}
	\end{subfigure}
    \caption{The probability of a candidate hypervelocity star being bound to the Galaxy versus the difference between the Galactocentric rest-frame velocity and the escape speed. The error bars incorporate errors and correlations in the distances, radial velocities and proper motions of the stars, as well as the uncertainties in the Solar kinematics and the Milky Way escape velocity (see the Appendix for more detail). Some stars are missing either the radial velocity or proper motions and thus the bound probability is only an upper limit (these objects are indicated with a triangle). The size of the point reflects whether the star is a giant (large), dwarf (medium) or a white dwarf or subdwarf (small).}
    \label{fig:pbound}
\end{figure*}

\begin{figure}
	\includegraphics[width=\linewidth,trim={0cm 0cm 1cm 1.3cm},clip]{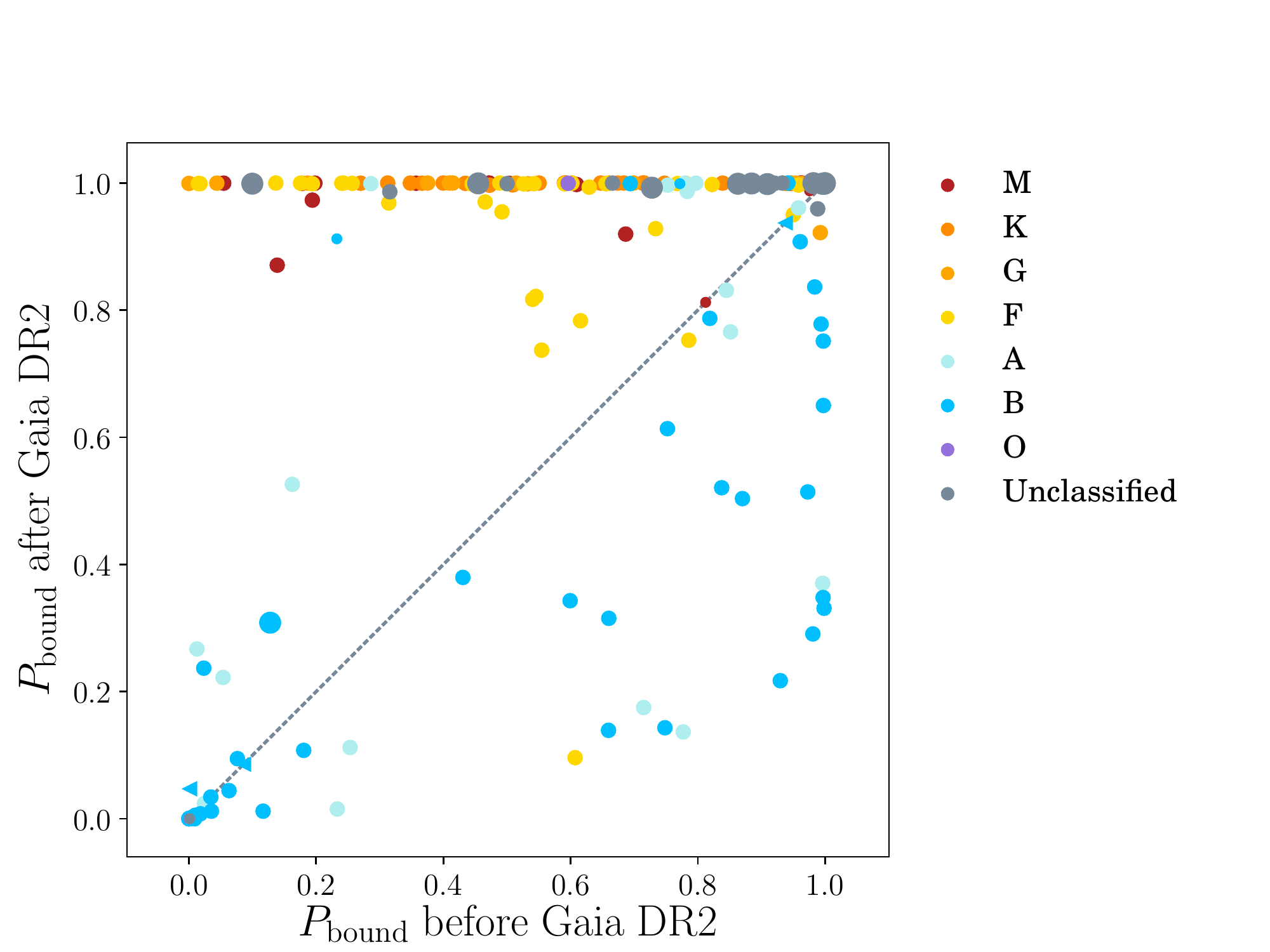}
    \caption{Probability $P_{\rm bound}$ that each high-velocity candidate is bound to the Milky Way before ($x$-axis) and after ($y$-axis) the inclusion of DR2. The shapes of the markers is as in Fig. \ref{fig:pbound}.}
    \label{fig:pboundpreafter}
\end{figure}

\footnotetext{\url{https://faststars.space/sky-locations/}}

\section{Results}
\label{sec:results}

\begin{figure}
\includegraphics[width=\linewidth]{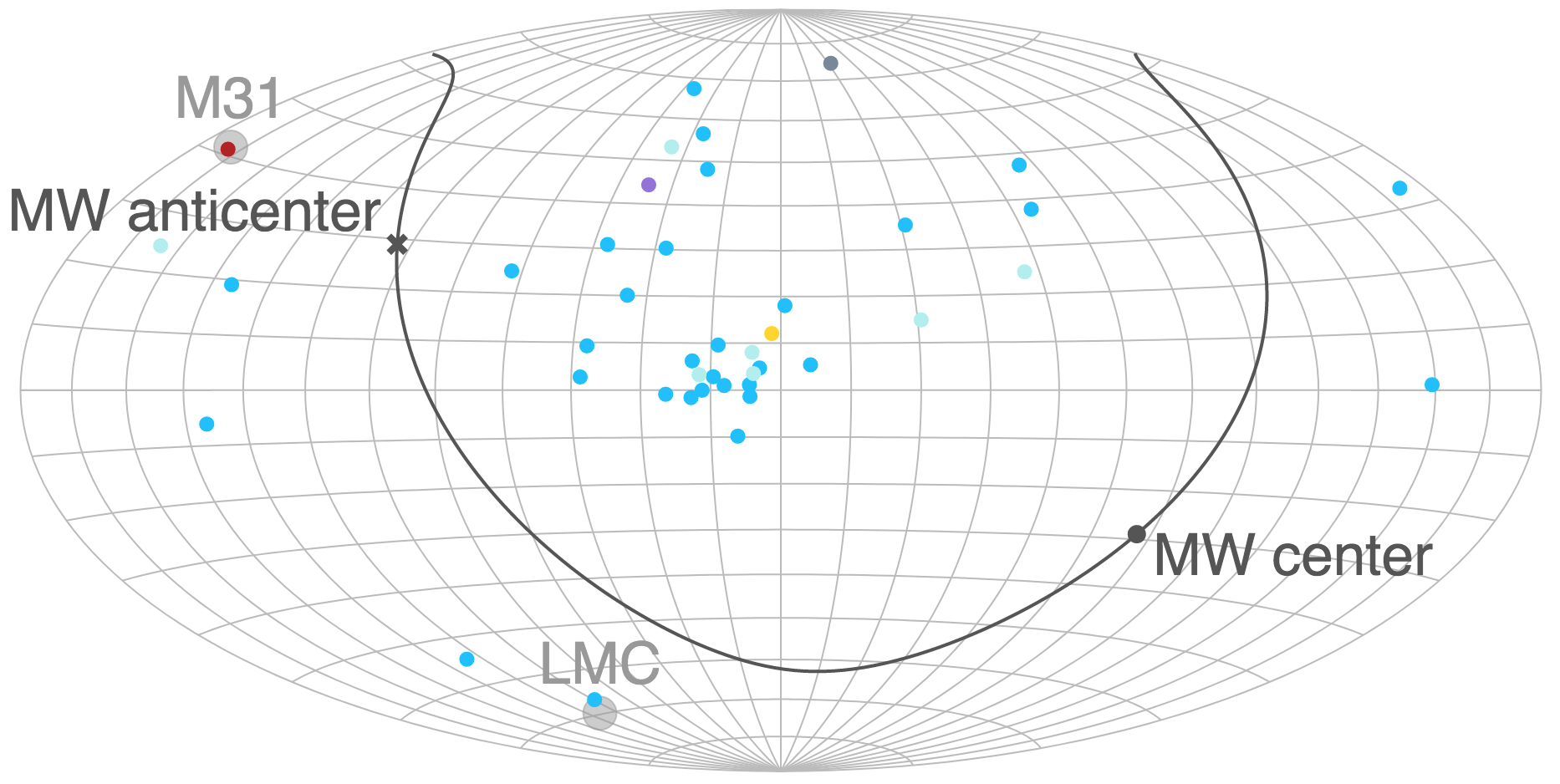}
    \caption{Same as Figure~\ref{fig:skymap}, but for hypervelocity candidates with $P_{\mathrm{bound}}<0.5$ (see Tab.~\ref{tab:candidates}).}
    \label{fig:unboundmap}
\end{figure}

The Open Fast Stars Catalog (presented in detail in Appendix \ref{sec:open}) automatically queries Gaia DR2 and calculates the posterior probability that each star is bound $P_{\mathrm{bound}}$ (the method is described in detail in the Appendix). Of the 524 candidate hypervelocity stars in our catalogue, 514 have Gaia photometry and 501 have Gaia astrometry in DR2. This compares with only 472 having photometry and 18 having astrometry in Gaia DR1. Almost all hypervelocity candidates now have precise proper motions and parallaxes, which transforms the landscape of hypervelocity star research. Before Gaia DR2, there were 71 candidates with $P_{\mathrm{bound}}<0.5$ and 132 candidates with $P_{\mathrm{bound}}>0.5$. After Gaia DR2, these numbers dramatically changed with 41 candidates with $P_{\mathrm{bound}}<0.5$ and 464 candidates with $P_{\mathrm{bound}}>0.5$. The increase in the numbers of classified stars is because Gaia provides parallaxes for the 321 candidates proposed by \citet{vickers_red_2015}, who had calculated photometric distances but not published them. We note that 428 hypervelocity candidates have a probability greater than 99\% of being bound to Milky Way and thus are ruled out. A caveat is that 5 stars are missing a radial velocity or proper motion in our catalogue and thus the $P_{\mathrm{bound}}$ is only an upper limit. It is possible that as the Open Fast Stars Catalog becomes more complete some of these candidates could be resurrected. A further caveat is that in this further analysis we do not consider the red giant J004330.06+405258.4, which is thought to be a hypervelocity star of M31 and is at a distance of $760\;\mathrm{kpc}$ \citep{evans_runaway_2015}. J004330.06+405258.4 is shown in Figs. \ref{fig:skymap} and \ref{fig:unboundmap} for completeness. We also do not consider Li2, the second candidate of \citet{li_19_2015}, because there are two radial velocity measurements in the literature which disagree: LAMOST reports $-60\pm10\;\mathrm{km}\;\mathrm{s}^{-1}$ whilst SDSS reports $-160.8\pm3.4\;\mathrm{km}\;\mathrm{s}^{-1}$. The simplest explanation is that this star is an unresolved binary and thus the reported radial velocities are not representative of the true systemic velocity.

In Figure \ref{fig:pbound}, we show the bound probability versus the difference between the Galactocentric rest-frame velocity and the escape velocity as a function of the spectral type. The overarching trend is for late-type FGKM stars to be assessed as more bound after DR2, while early-type OBA stars become less bound. This trend is made obvious in Fig. \ref{fig:pboundpreafter} where we directly compare $P_{\mathrm{bound}}$ computed before and after Gaia DR2; almost all the late-type stars are conclusively bound with DR2, whilst a large number of OBA stars have an increased probability of being unbound (they move to the lower right of this figure).

In Tab. \ref{tab:candidates}, we list all candidates which have $P_{\mathrm{bound}}<0.5$. This list of candidates comprises 38 B/A dwarfs, one subdwarf O star, one F9 dwarf and one white dwarf, and we will consider each of these categories in turn.

\begin{table}
\centering
\caption{The hypervelocity candidates with $P_{\mathrm{bound}}<0.5$ subdivided by original discovery survey or paper. The Hypervelocity Star Survey \citep{brown_discovery_2005,brown_hypervelocity_2007,brown_mmt_2014} has remarkably discovered 32 of these 41 stars, while the LAMOST HVS Survey \citep{zheng_first_2014,huang_discovery_2017} contributes a further 3.}
\label{tab:candidates}
\begin{tabular}{lll}
\hline Survey                    & \# & Names                          \\ \hline
Hypervelocity Star Survey & 32                                             & HVS1,4-10,12-24 and others \\
\citet{hirsch_us_2005}            & 1                                              & US708 (a.k.a. HVS2)            \\
\citet{edelmann_he_2005}           & 1                                              & HE 0437-5439 (a.k.a. HVS3)     \\
\citet{heber_b-type_2008}              & 1                                              & HD 271791                      \\
\citet{tillich_hypermuchfuss_2009}           & 1                                              & SDSS J013655.91+242546.0       \\
LAMOST HVS Survey         & 3                                              & LAMOST-HVS1-3                  \\
\citet{li_19_2015}                 & 1                                              & Li10 (F9 dwarf)                           \\
\citet{vennes_unusual_2017}            & 1                                              & GD 492 (white dwarf)    \\ \hline
\end{tabular}
\end{table}

\subsection{The early-type B/A candidates}
\label{sec:ba}

Over the past 13 years the Hypervelocity Star Survey \citep{brown_discovery_2005,brown_hypervelocity_2007,brown_mmt_2014} has discovered many tens of faint, blue stars in the halo of the Milky Way. These stars were classified as hypervelocity stars based solely on their large radial velocities and thus could not be ruled out by Gaia astrometry; by measuring their proper motions Gaia was only increasing their Galactocentric rest-frame velocity. This argument extends to most of the early-type stars shown in Fig. \ref{fig:pbound}, except for close stars such as HD 271791 at $21\pm4\;\mathrm{kpc}$ \citep{heber_b-type_2008} who had previously measured proper motions, and thus explains their trend to being more likely unbound.

\begin{figure}
\includegraphics[width=\linewidth]{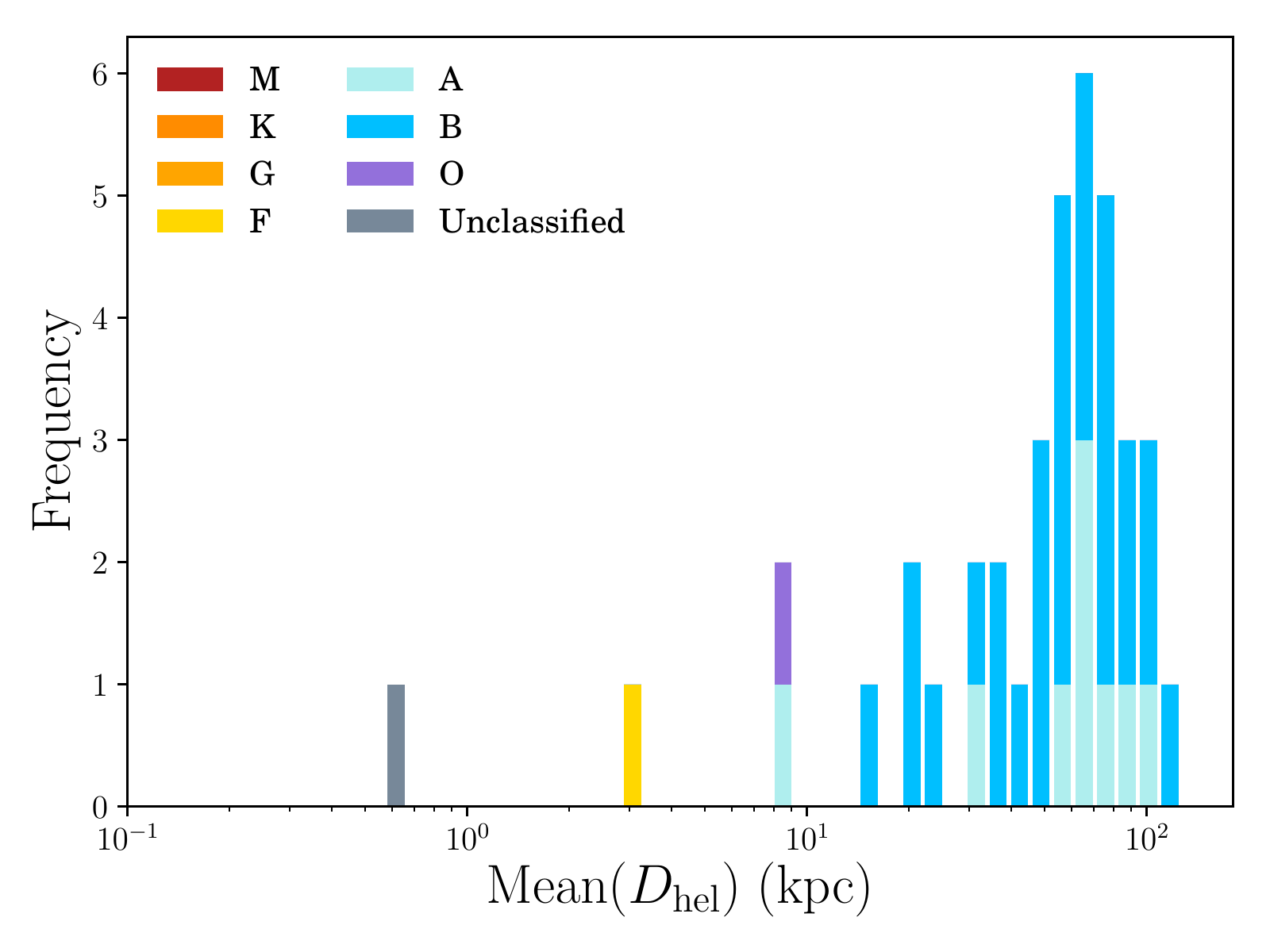}
    \caption{The heliocentric distance distribution of hypervelocity candidates with $P_{\mathrm{bound}}<0.5$ (see Tab.~\ref{tab:candidates}).}
    \label{fig:distances}
\end{figure}

As the majority of the remaining hypervelocity candidates are early-type, the distance distribution (see Fig. \ref{fig:distances}) of hypervelocity stars is now dominated by objects in the distance range ${10-110}\;\mathrm{kpc}$, with a modal distance of around $70\;\mathrm{kpc}$. The mean hypervelocity candidate with $P_{\mathrm{bound}}<0.5$ is now more distant than the LMC
\citep[$49.97\pm1.126\;\mathrm{kpc}$,][]{2013Natur.495...76P}. A further consequence is that the sky distribution is no longer homogeneous within the Northern equatorial hemisphere. The clump near the centre of the plot is the well-known clustering of early-type hypervelocity stars near the Leo constellation \citep[e.g.][]{brown_anisotropic_2009}. The star located beside the LMC is HVS3: an $8\;\mathrm{M}_{\odot}$ star thought to have been ejected from the LMC \citep{edelmann_he_2005,gualandris_hypervelocity_2007,2018arXiv180410197E}.

We note that the distance distribution shown in Fig. \ref{fig:distances} is biased by the way that the early-type hypervelocity stars were discovered. The Hypervelocity Star Survey \citep{brown_discovery_2005,brown_hypervelocity_2007,brown_mmt_2014} selected for blue, faint objects at high latitudes, because a B type star would require a large velocity to reach the halo within its lifetime. Thus by construction our sample of hypervelocity stars is biased towards stars at great distances. The existence or non-existence of early-type hypervelocity stars closer to the Galaxy will allow us to tell whether the hypervelocity stars have a Galactic or extragalactic origin.

\subsection{Type Ia supernova donors and survivors}
\label{sec:Ia}
Both US708 \citep{hirsch_us_2005} and GD 492 \citep{vennes_unusual_2017} are thought to be associated with Supernova Ia. We discuss each briefly. Note that the other white dwarf hypervelocity candidate SDSSJ124043.01+671034.68 \citep{2016Sci...352...67K} is confirmed with Gaia DR2 to be bound to the Galaxy.

\citet{hirsch_us_2005} initially conjectured that US708, a helium subdwarf O star, was formed in the merger of two helium white dwarfs during an interaction with the SMBH at the Galactic centre. However, \citet{justham_type_2009} proposed that this star was more consistent with having been the low-mass helium donor to a massive white dwarf and thence having being ejected by the resulting thermonuclear supernova Ia, and a subsequent spectroscopic and kinematic analysis confirmed this as the likely origin channel \citep{geier_fastest_2015}.

\citet{vennes_unusual_2017} discovered the low-mass, high proper motion white dwarf GD 492 and found that it had an atmosphere rich with intermediate-elements. The conclusion reached by \citet{vennes_unusual_2017} was that GD 492 is the partially burnt remnant of a subluminous supernova Ia. \citet{raddi_anatomy_2018}  concurred with the remnant hypothesis for GD 492 and used Gaia DR2 astrometry to constrain the progenitor. In this scenario, it must have been in a short period binary ($30-60\;\mathrm{min}$) with a $0.8-1.32\;\mathrm{M}_{\odot}$ companion. \citet{raddi_anatomy_2018} note that Gaia DR2 astrometry confirms GD 492 as the closest hypervelocity star to the Sun ($d_{\mathrm{hel}}=632\pm14\;\mathrm{pc}$).

\subsection{The remaining late-type hypervelocity candidate}
\label{sec:li10}

\begin{figure*}
	\includegraphics[width=\linewidth]{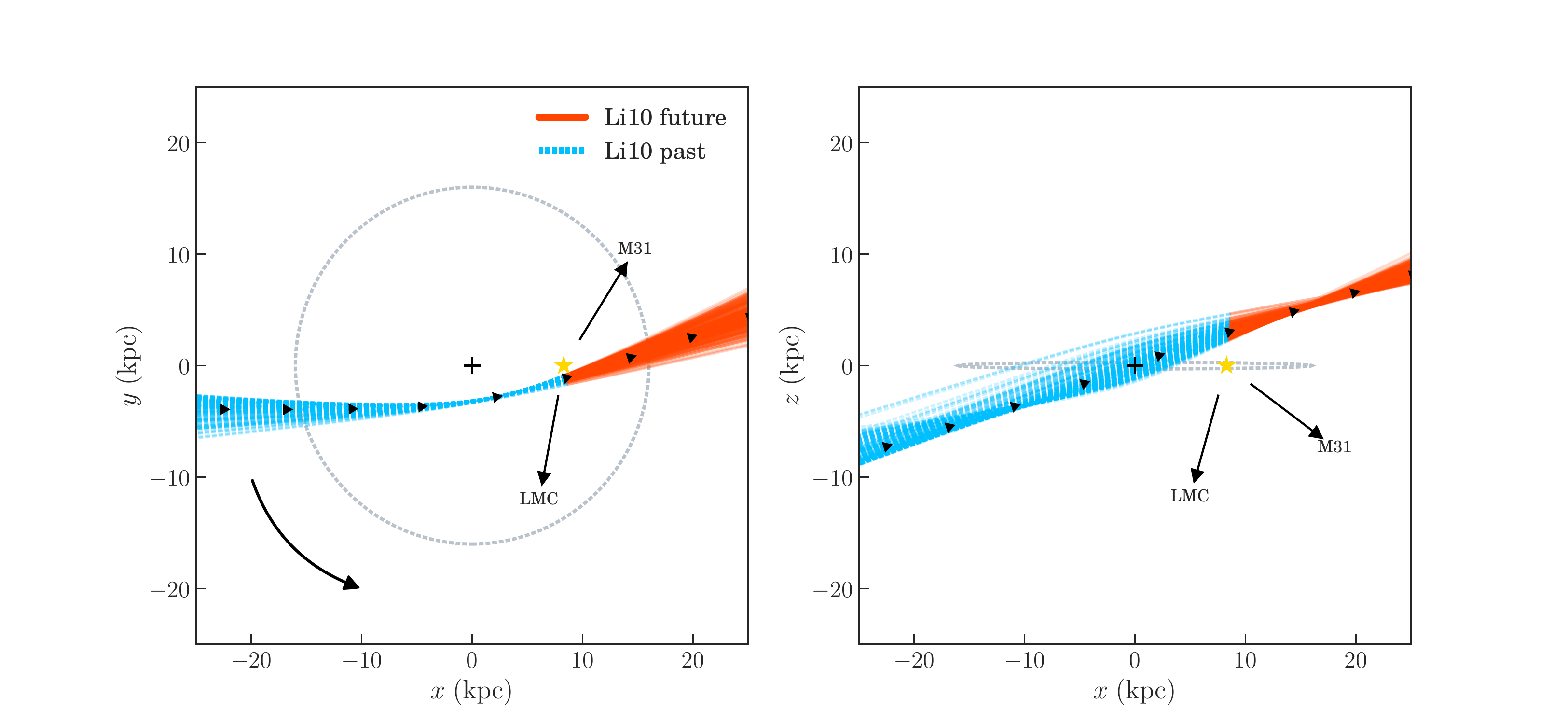}
    \caption{Past (dashed components of each curve) and future (solid components of each curve) realizations of the trajectory of the candidate high-velocity late-type star LAMOST J115209.12+120258.0. The size of the Milky Way's thin disk, assumed to be 32~kpc in diameter and 0.6~kpc in height, is shown by the dashed gray contours. Arrows pointing in the directions of M31 and the LMC are labeled. Orbits were calculated in the MWPotential2014 potential using the \textsc{Python} Galactic dynamics framework \textsc{Galpy} \citep{bovy_galpy:_2015}. The rotation direction of the Milky Way disc is indicated by the long arrow and the short arrows indicate time steps of $10\;\mathrm{Myr}$ along the orbit.}
    \label{fig:orbits}
\end{figure*}

Li10 was one of 19 candidates proposed by \citet{li_19_2015} based on LAMOST spectroscopy and proper motions from SDSS+USNO-B, and was found to have a 50\% probability of being bound in the \citet{2008ApJ...684.1143X} potential. Li10 was not directly discussed in \citet{li_19_2015} and does not appear to have been discussed elsewhere in the literature. As shown in Fig. \ref{fig:orbits}, the trajectory of this star back in time shows it passing within a few kiloparsecs of the Galactic centre. However, the pericentric radius is well constrained to be $3.3\pm 0.2\;\mathrm{kpc}$ and thus the Hills mechanism is ruled out as a possible explanation. One possibility is that the star is a runaway star that was either dynamically ejected from a star cluster or kicked by the supernova of a much more massive companion. \citet{tauris_maximum_2015} found that kicks of up to $1280\;\mathrm{km}\;\mathrm{s}^{-1}$ were possible in the supernova scenario for G/K dwarfs, which is much greater than the Galactocentric rest-frame velocity $643\pm93\;\mathrm{km}\;\mathrm{s}^{-1}$ of Li10. However, such velocities are expected to be extremely rare.

Li10 is consistent with having passed through the disc roughly $15\;\mathrm{Myr}$ ago, however we note that this is only a small fraction of the main sequence lifetime of an F9 star and thus we cannot use this time as an estimator of the flight time. The possibility of hypervelocity stars arriving in the Milky Way from M31 \citep{sherwin_hypervelocity_2008} or the LMC \citep{boubert_dipole_2016,boubert_hypervelocity_2017} has been suggested in the literature. However, the orbit of this star is not aligned with either of these galaxies (see Fig. \ref{fig:orbits}).  If the star were to turn out to be bound after later Gaia data releases, then the natural interpretation is that this star is a fast-moving denizen of the halo on an extremely radial orbit. An alternative possibility is that Li10 has a binary companion and thus that the radial velocity from LAMOST DR1 has a large contribution from the binary orbital motion. There is no source in Gaia DR2 within $45\;\mathrm{arcsec}$ of Li10 and thus any companion would need to be either a low-mass dwarf or a compact object (likely a white dwarf or neutron star).

To test the close binary hypothesis we obtained a spectrum of LAMOST J115209.12+120258.0 with the Goodman Spectrograph \citep{2004SPIE.5492..331C} on the SOAR telescope on UT 2018 Apr 29. We used a 0.95\arcsec slit and a 1200 l mm$^{-1}$ grating, giving a spectral resolution of about 1.7\AA. A single 1800-sec exposure was obtained. The spectrum was reduced and optimally extracted in the usual manner. We determined the barycentric radial velocity of the star through cross-correlation with a template of similar spectral type taken with the same setup, finding a value of $v_r = 234\pm5\;\mathrm{km}\;\mathrm{s}^{-1}$ which we use throughout this work. The LAMOST DR1 velocity of this star is listed as $206\pm15\;\mathrm{km}\;\mathrm{s}^{-1}$, which is marginally consistent with the new measurement. To check this, we downloaded the LAMOST spectrum and re-derived the radial velocity through cross-correlation. In the region of the Mg$b$ line we reproduce the published velocity; if instead we use the Ca triplet region, the LAMOST velocity is $223\pm5$ km s$^{-1}$, (random uncertainty only), which is closer to the SOAR/Goodman value. We conclude there is no significant evidence for a radial velocity shift between these two spectra and hence no evidence that this star is in a close binary. Thus, Li10 appears to be the only known late-type hypervelocity star.

\section{Conclusions}
In this paper, we have combined the historical data on high-velocity stars with data from Gaia's second data release. There is a single candidate late-type high-velocity object (LAMOST J115209.12+120258.0) that has a reasonably high probability of being unbound from the Milky Way and thus hypervelocity. However, the overwhelming majority of the historical late-type high-velocity candidates are almost certainly bound to the Milky Way. This is a clear demonstration of the superiority of space-based astrometry from Gaia over the earlier ground-based proper motion catalogues. It is anticipated that further Gaia DR2 studies will reveal many late-type high-velocity candidates, which will be added to the Open Fast Stars Catalog when they are announced.

\section*{Acknowledgements}

We thank Warren~R.~Brown, Saurabh~W.~Jha, Ken~J.~Shen, and Angus~Williams for valuable input. D. Boubert thanks the UK Science and Technology Facilities Council for supporting his PhD. K. Hawkins thanks the Simons Society of Fellows and the Flatiron Institute Center for Computational Astrophysics. I. Ginsburg was supported in part by Harvard University funds and the Institute for Theory and Computation. J.S. acknowledges support from the Packard Foundation. This work has made use of data from the European Space Agency (ESA) mission {\it Gaia}(\url{https://www.cosmos.esa.int/gaia}), processed by the {\it Gaia} Data Processing and Analysis Consortium (DPAC, \url{https://www.cosmos.esa.int/web/gaia/dpac/consortium}). Funding for the DPAC has been provided by national institutions, in particular the institutions participating in the {\it Gaia} Multilateral Agreement. Based on observations obtained at the Southern Astrophysical Research (SOAR) telescope, which is a joint project of the Minist\'{e}rio da Ci\^{e}ncia, Tecnologia, Inova\c{c}\~{o}es e Comunica\c{c}\~{o}es (MCTIC) do Brasil, the U.S. National Optical Astronomy Observatory (NOAO), the University of North Carolina at Chapel Hill (UNC), and Michigan State University (MSU).




\bibliographystyle{mnras}
\bibliography{references,wyn}

\begin{thebibliography}{}
\makeatletter
\relax
\def\mn@urlcharsother{\let\do\@makeother \do\$\do\&\do\#\do\^\do\_\do\%\do\~}
\def\mn@doi{\begingroup\mn@urlcharsother \@ifnextchar [ {\mn@doi@}
  {\mn@doi@[]}}
\def\mn@doi@[#1]#2{\def\@tempa{#1}\ifx\@tempa\@empty \href
  {http://dx.doi.org/#2} {doi:#2}\else \href {http://dx.doi.org/#2} {#1}\fi
  \endgroup}
\def\mn@eprint#1#2{\mn@eprint@#1:#2::\@nil}
\def\mn@eprint@arXiv#1{\href {http://arxiv.org/abs/#1} {{\tt arXiv:#1}}}
\def\mn@eprint@dblp#1{\href {http://dblp.uni-trier.de/rec/bibtex/#1.xml}
  {dblp:#1}}
\def\mn@eprint@#1:#2:#3:#4\@nil{\def\@tempa {#1}\def\@tempb {#2}\def\@tempc
  {#3}\ifx \@tempc \@empty \let \@tempc \@tempb \let \@tempb \@tempa \fi \ifx
  \@tempb \@empty \def\@tempb {arXiv}\fi \@ifundefined
  {mn@eprint@\@tempb}{\@tempb:\@tempc}{\expandafter \expandafter \csname
  mn@eprint@\@tempb\endcsname \expandafter{\@tempc}}}

\bibitem[\protect\citeauthoryear{{Abadi}, {Navarro}  \& {Steinmetz}}{{Abadi}
  et~al.}{2009}]{Ab09}
{Abadi} M.~G.,  {Navarro} J.~F.,   {Steinmetz} M.,  2009, \mn@doi [\apjl]
  {10.1088/0004-637X/691/2/L63}, \href
  {http://adsabs.harvard.edu/abs/2009ApJ...691L..63A} {691, L63}

\bibitem[\protect\citeauthoryear{{Abolfathi} et~al.,}{{Abolfathi}
  et~al.}{2018}]{2018ApJS..235...42A}
{Abolfathi} B.,  et~al., 2018, \mn@doi [\apjs] {10.3847/1538-4365/aa9e8a},
  \href {http://adsabs.harvard.edu/abs/2018ApJS..235...42A} {235, 42}

\bibitem[\protect\citeauthoryear{Astraatmadja \& Bailer-Jones}{Astraatmadja \&
  Bailer-Jones}{2016}]{astraatmadja_estimating_2016}
Astraatmadja T.~L.,  Bailer-Jones C. A.~L.,  2016, \mn@doi [ApJ]
  {10.3847/0004-637X/832/2/137}, 832, 137

\bibitem[\protect\citeauthoryear{Bailer-Jones}{Bailer-Jones}{2015}]{bailer-jones_estimating_2015}
Bailer-Jones C. A.~L.,  2015, \mn@doi [PASP] {10.1086/683116}, 127, 994

\bibitem[\protect\citeauthoryear{{Bailer-Jones}, {Rybizki}, {Fouesneau},
  {Mantelet}  \& {Andrae}}{{Bailer-Jones} et~al.}{2018}]{2018arXiv180410121B}
{Bailer-Jones} C.~A.~L.,  {Rybizki} J.,  {Fouesneau} M.,  {Mantelet} G.,
  {Andrae} R.,  2018, preprint, \href
  {https://ui.adsabs.harvard.edu/#abs/2018arXiv180410121B} {p.
  arXiv:1804.10121} (\mn@eprint {arXiv} {1804.10121})

\bibitem[\protect\citeauthoryear{Boubert \& Evans}{Boubert \&
  Evans}{2016}]{boubert_dipole_2016}
Boubert D.,  Evans N.~W.,  2016, \mn@doi [ApJ] {10.3847/2041-8205/825/1/L6},
  825, L6

\bibitem[\protect\citeauthoryear{Boubert, Erkal, Evans  \& Izzard}{Boubert
  et~al.}{2017}]{boubert_hypervelocity_2017}
Boubert D.,  Erkal D.,  Evans N.~W.,   Izzard R.~G.,  2017, \mn@doi [MNRAS]
  {10.1093/mnras/stx848}, 469, 2151

\bibitem[\protect\citeauthoryear{Bovy}{Bovy}{2015}]{bovy_galpy:_2015}
Bovy J.,  2015, \mn@doi [ApJS] {10.1088/0067-0049/216/2/29}, 216, 29

\bibitem[\protect\citeauthoryear{Brown}{Brown}{2015}]{brown_hypervelocity_2015}
Brown W.~R.,  2015, \mn@doi [ARA\&A] {10.1146/annurev-astro-082214-122230}, 53,
  15

\bibitem[\protect\citeauthoryear{Brown, Cai  \& DasGupta}{Brown
  et~al.}{2001}]{brown_interval_2001}
Brown L.~D.,  Cai T.~T.,   DasGupta A.,  2001, \mn@doi [Statist. Sci.]
  {10.1214/ss/1009213286}, 16, 101

\bibitem[\protect\citeauthoryear{Brown, Geller, Kenyon  \& Kurtz}{Brown
  et~al.}{2005}]{brown_discovery_2005}
Brown W.~R.,  Geller M.~J.,  Kenyon S.~J.,   Kurtz M.~J.,  2005, \mn@doi [ApJ]
  {10.1086/429378}, 622, L33

\bibitem[\protect\citeauthoryear{Brown, Geller, Kenyon  \& Kurtz}{Brown
  et~al.}{2006}]{brown_hypervelocity_2006}
Brown W.~R.,  Geller M.~J.,  Kenyon S.~J.,   Kurtz M.~J.,  2006, \mn@doi [ApJ]
  {10.1086/505165}, 647, 303

\bibitem[\protect\citeauthoryear{Brown, Geller, Kenyon, Kurtz  \&
  Bromley}{Brown et~al.}{2007}]{brown_hypervelocity_2007}
Brown W.~R.,  Geller M.~J.,  Kenyon S.~J.,  Kurtz M.~J.,   Bromley B.~C.,
  2007, \mn@doi [ApJ] {10.1086/523642}, 671, 1708

\bibitem[\protect\citeauthoryear{Brown, Geller, Kenyon  \& Bromley}{Brown
  et~al.}{2009}]{brown_anisotropic_2009}
Brown W.~R.,  Geller M.~J.,  Kenyon S.~J.,   Bromley B.~C.,  2009, \mn@doi
  [ApJ] {10.1088/0004-637X/690/1/L69}, 690, L69

\bibitem[\protect\citeauthoryear{Brown, Geller  \& Kenyon}{Brown
  et~al.}{2014}]{brown_mmt_2014}
Brown W.~R.,  Geller M.~J.,   Kenyon S.~J.,  2014, \mn@doi [ApJ]
  {10.1088/0004-637X/787/1/89}, 787, 89

\bibitem[\protect\citeauthoryear{{Clemens}, {Crain}  \& {Anderson}}{{Clemens}
  et~al.}{2004}]{2004SPIE.5492..331C}
{Clemens} J.~C.,  {Crain} J.~A.,   {Anderson} R.,  2004, in {Moorwood}
  A.~F.~M.,  {Iye} M.,  eds,  \procspie Vol. 5492, Ground-based Instrumentation
  for Astronomy. pp 331--340, \mn@doi{10.1117/12.550069}

\bibitem[\protect\citeauthoryear{{Cui} et~al.,}{{Cui} et~al.}{2012}]{LAMOST}
{Cui} X.-Q.,  et~al., 2012, \mn@doi [Research in Astronomy and Astrophysics]
  {10.1088/1674-4527/12/9/003}, \href
  {http://adsabs.harvard.edu/abs/2012RAA....12.1197C} {12, 1197}

\bibitem[\protect\citeauthoryear{Edelmann, Napiwotzki, Heber, Christlieb  \&
  Reimers}{Edelmann et~al.}{2005}]{edelmann_he_2005}
Edelmann H.,  Napiwotzki R.,  Heber U.,  Christlieb N.,   Reimers D.,  2005,
  \mn@doi [ApJ] {10.1086/498940}, 634, L181

\bibitem[\protect\citeauthoryear{{Erkal}, {Boubert}, {Gualandris}, {Evans}  \&
  {Antonini}}{{Erkal} et~al.}{2018}]{2018arXiv180410197E}
{Erkal} D.,  {Boubert} D.,  {Gualandris} A.,  {Evans} N.~W.,   {Antonini} F.,
  2018, preprint, \href {http://adsabs.harvard.edu/abs/2018arXiv180410197E} {}
  (\mn@eprint {arXiv} {1804.10197})

\bibitem[\protect\citeauthoryear{Evans \& Massey}{Evans \&
  Massey}{2015}]{evans_runaway_2015}
Evans K.~A.,  Massey P.,  2015, \mn@doi [AJ] {10.1088/0004-6256/150/5/149},
  150, 149

\bibitem[\protect\citeauthoryear{{Foreman-Mackey}, {Hogg}, {Lang}  \&
  {Goodman}}{{Foreman-Mackey} et~al.}{2013}]{2013PASP..125..306F}
{Foreman-Mackey} D.,  {Hogg} D.~W.,  {Lang} D.,   {Goodman} J.,  2013, \mn@doi
  [\pasp] {10.1086/670067}, \href
  {http://adsabs.harvard.edu/abs/2013PASP..125..306F} {125, 306}

\bibitem[\protect\citeauthoryear{{Gaia Collaboration} et~al.,}{{Gaia
  Collaboration} et~al.}{2016}]{gaia_collaboration_gaia_2016}
{Gaia Collaboration} et~al., 2016, \mn@doi [A\&A]
  {10.1051/0004-6361/201629512}, 595, A2

\bibitem[\protect\citeauthoryear{{Gaia Collaboration}, {Brown}, {Vallenari},
  {Prusti}, {de Bruijne}, {Babusiaux}  \& {Bailer-Jones}}{{Gaia Collaboration}
  et~al.}{2018}]{2018arXiv180409365G}
{Gaia Collaboration} {Brown} A.~G.~A.,  {Vallenari} A.,  {Prusti} T.,  {de
  Bruijne} J.~H.~J.,  {Babusiaux} C.,   {Bailer-Jones} C.~A.~L.,  2018,
  preprint, \href {http://adsabs.harvard.edu/abs/2018arXiv180409365G} {}
  (\mn@eprint {arXiv} {1804.09365})

\bibitem[\protect\citeauthoryear{Geier et~al.,}{Geier
  et~al.}{2015}]{geier_fastest_2015}
Geier S.,  et~al., 2015, \mn@doi [Science] {10.1126/science.1259063}, 347, 1126

\bibitem[\protect\citeauthoryear{Ginsburg \& Loeb}{Ginsburg \&
  Loeb}{2007}]{ginsburg_hypervelocity_2007}
Ginsburg I.,  Loeb A.,  2007, \mn@doi [MNRAS]
  {10.1111/j.1365-2966.2007.11461.x}, 376, 492

\bibitem[\protect\citeauthoryear{Green et~al.,}{Green
  et~al.}{2015}]{green_three-dimensional_2015}
Green G.~M.,  et~al., 2015, \mn@doi [ApJ] {10.1088/0004-637X/810/1/25}, 810, 25

\bibitem[\protect\citeauthoryear{Green et~al.,}{Green
  et~al.}{2018}]{green_galactic_2018}
Green G.~M.,  et~al., 2018, preprint, 1801, arXiv:1801.03555

\bibitem[\protect\citeauthoryear{Gualandris \& Portegies~Zwart}{Gualandris \&
  Portegies~Zwart}{2007}]{gualandris_hypervelocity_2007}
Gualandris A.,  Portegies~Zwart S.,  2007, \mn@doi [MNRAS]
  {10.1111/j.1745-3933.2007.00280.x}, 376, L29

\bibitem[\protect\citeauthoryear{Guillochon, Parrent, Kelley  \&
  Margutti}{Guillochon et~al.}{2017}]{guillochon_open_2017}
Guillochon J.,  Parrent J.,  Kelley L.~Z.,   Margutti R.,  2017, \mn@doi [ApJ]
  {10.3847/1538-4357/835/1/64}, 835, 64

\bibitem[\protect\citeauthoryear{Heber, Edelmann, Napiwotzki, Altmann  \&
  Scholz}{Heber et~al.}{2008}]{heber_b-type_2008}
Heber U.,  Edelmann H.,  Napiwotzki R.,  Altmann M.,   Scholz R.-D.,  2008,
  \mn@doi [A\&A] {10.1051/0004-6361:200809767}, 483, L21

\bibitem[\protect\citeauthoryear{Hills}{Hills}{1988}]{hills_hyper-velocity_1988}
Hills J.~G.,  1988, \mn@doi [Nature] {10.1038/331687a0}, 331, 687

\bibitem[\protect\citeauthoryear{Hirsch, Heber, O'Toole  \& Bresolin}{Hirsch
  et~al.}{2005}]{hirsch_us_2005}
Hirsch H.~A.,  Heber U.,  O'Toole S.~J.,   Bresolin F.,  2005, \mn@doi [A\&A]
  {10.1051/0004-6361:200500212}, 444, L61

\bibitem[\protect\citeauthoryear{Huang et~al.,}{Huang
  et~al.}{2017}]{huang_discovery_2017}
Huang Y.,  et~al., 2017, preprint, 1708, arXiv:1708.08602

\bibitem[\protect\citeauthoryear{Johnson \& Soderblom}{Johnson \&
  Soderblom}{1987}]{johnson_calculating_1987}
Johnson D. R.~H.,  Soderblom D.~R.,  1987, \mn@doi [ApJ] {10.1086/114370}, 93,
  864

\bibitem[\protect\citeauthoryear{Justham, Wolf, Podsiadlowski  \& Han}{Justham
  et~al.}{2009}]{justham_type_2009}
Justham S.,  Wolf C.,  Podsiadlowski P.,   Han Z.,  2009, \mn@doi [Astronomy
  and Astrophysics] {10.1051/0004-6361:200810106}, 493, 1081

\bibitem[\protect\citeauthoryear{{Kepler}, {Koester}  \& {Ourique}}{{Kepler}
  et~al.}{2016}]{2016Sci...352...67K}
{Kepler} S.~O.,  {Koester} D.,   {Ourique} G.,  2016, \mn@doi [Science]
  {10.1126/science.aad6705}, \href
  {http://adsabs.harvard.edu/abs/2016Sci...352...67K} {352, 67}

\bibitem[\protect\citeauthoryear{{Li}, {Luo}, {Zhao}, {Lu}, {Ren}  \&
  {Zuo}}{{Li} et~al.}{2012}]{Li12}
{Li} Y.,  {Luo} A.,  {Zhao} G.,  {Lu} Y.,  {Ren} J.,   {Zuo} F.,  2012, \mn@doi
  [\apjl] {10.1088/2041-8205/744/2/L24}, \href
  {http://adsabs.harvard.edu/abs/2012ApJ...744L..24L} {744, L24}

\bibitem[\protect\citeauthoryear{Li et~al.,}{Li et~al.}{2015}]{li_19_2015}
Li Y.-B.,  et~al., 2015, \mn@doi [RAA] {10.1088/1674-4527/15/8/018}, 15, 1364

\bibitem[\protect\citeauthoryear{{Luo} et~al.,}{{Luo} et~al.}{2016}]{Luo:2016a}
{Luo} A.-L.,  et~al., 2016, VizieR Online Data Catalog, 5149

\bibitem[\protect\citeauthoryear{{Luri} et~al.,}{{Luri}
  et~al.}{2018}]{2018arXiv180409376L}
{Luri} X.,  et~al., 2018, preprint, \href
  {http://adsabs.harvard.edu/abs/2018arXiv180409376L} {} (\mn@eprint {arXiv}
  {1804.09376})

\bibitem[\protect\citeauthoryear{Marchetti, Rossi, Kordopatis, Brown, Rimoldi,
  Starkenburg, Youakim  \& Ashley}{Marchetti
  et~al.}{2017}]{marchetti_artificial_2017}
Marchetti T.,  Rossi E.~M.,  Kordopatis G.,  Brown A. G.~A.,  Rimoldi A.,
  Starkenburg E.,  Youakim K.,   Ashley R.,  2017, preprint, 1704,
  arXiv:1704.07990

\bibitem[\protect\citeauthoryear{Munn et~al.,}{Munn
  et~al.}{2004}]{munn_improved_2004}
Munn J.~A.,  et~al., 2004, \mn@doi [AJ] {10.1086/383292}, 127, 3034

\bibitem[\protect\citeauthoryear{{Munn} et~al.,}{{Munn} et~al.}{2008}]{Mu08}
{Munn} J.~A.,  et~al., 2008, \mn@doi [\aj] {10.1088/0004-6256/136/2/895}, \href
  {http://adsabs.harvard.edu/abs/2008AJ....136..895M} {136, 895}

\bibitem[\protect\citeauthoryear{Palladino, Schlesinger, Holley-Bockelmann,
  Prieto, Beers, Lee  \& Schneider}{Palladino
  et~al.}{2014}]{palladino_hypervelocity_2014}
Palladino L.~E.,  Schlesinger K.~J.,  Holley-Bockelmann K.,  Prieto C.~A.,
  Beers T.~C.,  Lee Y.~S.,   Schneider D.~P.,  2014, \mn@doi [ApJ]
  {10.1088/0004-637X/780/1/7}, 780, 7

\bibitem[\protect\citeauthoryear{Pereira, Jilinski, Drake, de Castro, Ortega,
  Chavero  \& Roig}{Pereira et~al.}{2012}]{pereira_cd-621346:_2012}
Pereira C.~B.,  Jilinski E.,  Drake N.~A.,  de Castro D.~B.,  Ortega V.~G.,
  Chavero C.,   Roig F.,  2012, \mn@doi [A\&A] {10.1051/0004-6361/201219122},
  543, A58

\bibitem[\protect\citeauthoryear{{Pietrzy{\'n}ski} et~al.,}{{Pietrzy{\'n}ski}
  et~al.}{2013}]{2013Natur.495...76P}
{Pietrzy{\'n}ski} G.,  et~al., 2013, \mn@doi [\nat] {10.1038/nature11878},
  \href {http://adsabs.harvard.edu/abs/2013Natur.495...76P} {495, 76}

\bibitem[\protect\citeauthoryear{Raddi, Hollands, Gaensicke, Townsley, Hermes,
  Fusillo  \& Koester}{Raddi et~al.}{2018}]{raddi_anatomy_2018}
Raddi R.,  Hollands M.~A.,  Gaensicke B.~T.,  Townsley D.~M.,  Hermes J.~J.,
  Fusillo N. P.~G.,   Koester D.,  2018, arXiv:1804.09677 [astro-ph]

\bibitem[\protect\citeauthoryear{{Rybizki}, {Demleitner}, {Fouesneau},
  {Bailer-Jones}, {Rix}  \& {Andrae}}{{Rybizki}
  et~al.}{2018}]{2018arXiv180401427R}
{Rybizki} J.,  {Demleitner} M.,  {Fouesneau} M.,  {Bailer-Jones} C.,  {Rix}
  H.-W.,   {Andrae} R.,  2018, preprint, \href
  {http://adsabs.harvard.edu/abs/2018arXiv180401427R} {} (\mn@eprint {arXiv}
  {1804.01427})

\bibitem[\protect\citeauthoryear{Schlegel, Finkbeiner  \& Davis}{Schlegel
  et~al.}{1998}]{schlegel_maps_1998}
Schlegel D.~J.,  Finkbeiner D.~P.,   Davis M.,  1998, \mn@doi [ApJ]
  {10.1086/305772}, 500, 525

\bibitem[\protect\citeauthoryear{{Sch{\"o}nrich}}{{Sch{\"o}nrich}}{2012}]{2012MNRAS.427..274S}
{Sch{\"o}nrich} R.,  2012, \mn@doi [\mnras] {10.1111/j.1365-2966.2012.21631.x},
  \href {http://adsabs.harvard.edu/abs/2012MNRAS.427..274S} {427, 274}

\bibitem[\protect\citeauthoryear{{Sch{\"o}nrich}, {Binney}  \&
  {Dehnen}}{{Sch{\"o}nrich} et~al.}{2010}]{2010MNRAS.403.1829S}
{Sch{\"o}nrich} R.,  {Binney} J.,   {Dehnen} W.,  2010, \mn@doi [\mnras]
  {10.1111/j.1365-2966.2010.16253.x}, \href
  {http://adsabs.harvard.edu/abs/2010MNRAS.403.1829S} {403, 1829}

\bibitem[\protect\citeauthoryear{Sherwin, Loeb  \& O'Leary}{Sherwin
  et~al.}{2008}]{sherwin_hypervelocity_2008}
Sherwin B.~D.,  Loeb A.,   O'Leary R.~M.,  2008, \mn@doi [MNRAS]
  {10.1111/j.1365-2966.2008.13097.x}, 386, 1179

\bibitem[\protect\citeauthoryear{Tauris}{Tauris}{2015}]{tauris_maximum_2015}
Tauris T.~M.,  2015, \mn@doi [MNRAS] {10.1093/mnrasl/slu189}, 448, L6

\bibitem[\protect\citeauthoryear{Tillich et~al.,}{Tillich
  et~al.}{2009}]{tillich_hypermuchfuss_2009}
Tillich A.,  et~al., 2009, \mn@doi [Journal of Physics Conference Series]
  {10.1088/1742-6596/172/1/012009}, 172, 012009

\bibitem[\protect\citeauthoryear{Vennes, Nemeth, Kawka, Thorstensen, Khalack,
  Ferrario  \& Alper}{Vennes et~al.}{2017}]{vennes_unusual_2017}
Vennes S.,  Nemeth P.,  Kawka A.,  Thorstensen J.~R.,  Khalack V.,  Ferrario
  L.,   Alper E.~H.,  2017, preprint, 1708, arXiv:1708.05568

\bibitem[\protect\citeauthoryear{Vickers, Smith  \& Grebel}{Vickers
  et~al.}{2015}]{vickers_red_2015}
Vickers J.~J.,  Smith M.~C.,   Grebel E.~K.,  2015, \mn@doi [AJ]
  {10.1088/0004-6256/150/3/77}, 150, 77

\bibitem[\protect\citeauthoryear{Wenger et~al.,}{Wenger
  et~al.}{2000}]{wenger_simbad_2000}
Wenger M.,  et~al., 2000, \mn@doi [A\&AS] {10.1051/aas:2000332}, 143, 9

\bibitem[\protect\citeauthoryear{Williams, Belokurov, Casey  \& Evans}{Williams
  et~al.}{2017}]{williams_run:_2017}
Williams A.~A.,  Belokurov V.,  Casey A.~R.,   Evans N.~W.,  2017, preprint
  (arXiv:1701.01444)

\bibitem[\protect\citeauthoryear{{Xue} et~al.,}{{Xue}
  et~al.}{2008}]{2008ApJ...684.1143X}
{Xue} X.~X.,  et~al., 2008, \mn@doi [\apj] {10.1086/589500}, \href
  {http://adsabs.harvard.edu/abs/2008ApJ...684.1143X} {684, 1143}

\bibitem[\protect\citeauthoryear{{Yanny} et~al.,}{{Yanny} et~al.}{2009}]{SEGUE}
{Yanny} B.,  et~al., 2009, \mn@doi [\aj] {10.1088/0004-6256/137/5/4377}, \href
  {http://adsabs.harvard.edu/abs/2009AJ....137.4377Y} {137, 4377}

\bibitem[\protect\citeauthoryear{Zheng et~al.,}{Zheng
  et~al.}{2014}]{zheng_first_2014}
Zheng Z.,  et~al., 2014, \mn@doi [ApJ] {10.1088/2041-8205/785/2/L23}, 785, L23

\bibitem[\protect\citeauthoryear{Ziegerer, Volkert, Heber, Irrgang, Gänsicke
  \& Geier}{Ziegerer et~al.}{2015}]{ziegerer_candidate_2015}
Ziegerer E.,  Volkert M.,  Heber U.,  Irrgang A.,  Gänsicke B.~T.,   Geier S.,
   2015, \mn@doi [A\&A] {10.1051/0004-6361/201526052}, 576, L14

\makeatother
\end{thebibliography}



\appendix

\section{The Open Fast Stars Catalog}
\label{sec:open}

The papers which originally proposed the late-type hypervelocity candidates discussed in the main text often give measurements of properties not generically included in Gaia, such as spectral types, radial velocities and other spectroscopic parameters. Combining these properties in a systematic, rigorous fashion with Gaia astrometry and photometry is crucial to determining the nature of these candidates. To that end, we have created the Open Fast Stars Catalog (OFSC)\footnote{\url{https://faststars.space}} utilizing the AstroCats framework \citep{guillochon_open_2017}. The objective of the catalogue is to contain a curated collection of every measurement of all high-velocity star candidates in the literature, with each measurement having a citable origin, and to utilize the available data to provide additional value to the community interested in these objects. At present, the OFSC has targeted the data available for stars that may potentially be hypervelocity stars, however we plan to expand it to include pulsars, runaway stars, and halo stars in the near future.

Like the other Open Astronomy Catalogs\footnote{See \url{https://astrocats.space}}, the OFSC adds value to the existent data by providing derived quantities to the community. The catalog automatically computes the amount of extinction to each object (see Section~\ref{sec:othercatalogues}), velocities in various frames (heliocentric, galactocentric), observability at a user-specified time from various observatory locations, probability of boundedness to the Milky Way, and correlations between observed and derived quantities. The catalog also provides an interface for each object with a near-complete collection of its data. At the moment, the catalog only includes fast star spectroscopy from the SDSS survey \citep{2018ApJS..235...42A} and the LAMOST survey \citep{Luo:2016a}, as little is available from public repositories; we plan to collect this data from the community in the near future.

\subsection{Determination of boundedness}
The question of whether a star is hypervelocity can be more plainly phrased as ``is the total velocity $v_{\mathrm{grf}}$ in the Galactic rest-frame greater than the escape speed $v_{\mathrm{esc}}$ at its current location?'' Many of the papers which present hypervelocity star candidates give both the Galactic speed and the escape speed. However, the Galactic speed is sensitive to the assumed Solar position and peculiar motion and the escape speeds can vary by as much as $100\;\mathrm{km}\;\mathrm{s}^{-1}$ depending on the potential used. We therefore re-calculate the Galactic rest-frame speed and the escape speed for each candidate. Assuming that we have the equatorial position $(\alpha,\delta)$ and proper motions $(\mu_{\alpha\ast},\mu_{\delta})$ and the heliocentric distance $d$ and radial velocity $v_{\mathrm{r}}$, \citet{johnson_calculating_1987} provide formulae for obtaining the cylindrical Galactocentric position $(R,\theta,z)$ and velocity $(v_{\mathrm{R}},v_{\theta},v_{\mathrm{z}})$. The total Galactic rest-frame speed is then the the magnitude of this velocity. The escape velocity can be obtained from a fiducial escape velocity curve $v_{\mathrm{esc}}(r)$, for instance \citet{williams_run:_2017} parametrized this through
\begin{equation}
v_{\mathrm{esc}}(r) = v_{\mathrm{esc},\odot}\left(\frac{r}{R_{\odot}}\right)^{-\alpha/2},
\end{equation}
where $v_{\mathrm{esc},\odot}$ is the escape velocity at the position of the Sun, and obtained posterior constraints of $\alpha=0.37_{-0.09}^{+0.09}$ and $v_{\mathrm{esc},\odot}=521.26_{-30.23}^{+45.79}\;\mathrm{km}\;\mathrm{s}^{-1}$ using main-sequence turn-off, blue horizontal branch and K giant stars. We assume that the Milky Way disc rotates with a flat
circular velocity of
$V_{\mathrm{c}}=238\pm9\;\mathrm{km}\;\mathrm{s}^{-1}$ and that the
Sun orbits at the Galactocentric radius
$R_{\odot}=8.27\pm0.29\;\mathrm{kpc}$ with a peculiar velocity
$(U_{\odot}, V_{\odot},
W_{\odot})=(11.1\pm0.75\pm1,12.24\pm0.47\pm2,7.25\pm0.37\pm0.5)\;\mathrm{km}\;\mathrm{s}^{-1}$
\citep{2010MNRAS.403.1829S,2012MNRAS.427..274S}.

The method outlined in the previous paragraph would give a single $v_{\mathrm{grf}}$ and $v_{\mathrm{esc}}$ for each candidate and thus reduces the question of boundedness to simply which quantity is the greater. However, in practise, each of the heliocentric quantities will have attached uncertainties and it is vital to account for these. Care is required, because the uncertainty in the distance causes the uncertainties in the Galactic speed and escape speed to be correlated. The uncertainties in different measurements may themselves be correlated, with the notable example of Gaia providing the covariance matrix between the positions, parallax and proper motions. An additional complication is the need to use a sensible prior on the true distance of a star when converting parallax to distance \citep{bailer-jones_estimating_2015}. We assume the exponentially-decreasing volume prior of \citet{astraatmadja_estimating_2016} which is a Gamma distribution with shape parameter $k=3$ and scale parameter $L$. When applying this methodology to Gaia DR1 astrometry we used $L=1.35\;\mathrm{kpc}$ as recommended by \citet{astraatmadja_estimating_2016}. For Gaia DR2 astrometry we used the more complicated spatially-varying scale-length of \citet{2018arXiv180410121B} which was tuned to a mock of the contents of Gaia DR2 \citep{2018arXiv180401427R}. We assume that the uncertainty on the positions and velocities are adequately described by a multivariate normal distribution centred on the measured values and with covariance matrix $\boldsymbol{C}$, where the off-diagonal terms are zero unless the star has Gaia astrometry. For each star, we use \textsc{emcee} \citep{2013PASP..125..306F} to draw $10^4$ samples from the multivariate normal likelihood and distance prior. For any star with a photometric distance that has an uncertainty we replace the standard prior with a Gaussian distance prior centred on the photometric distance and a width equal to the uncertainty on that distance. We additionally sample in the Gaussian statistical and systematic uncertainty of the Solar position and motion. A further complication is that we do not yet have a firm knowledge of the escape velocity from the Milky Way, which we account for by sampling the parameters of the escape velocity curve from the posterior of \citet{williams_run:_2017}. The sampled values are then processed as described in the previous paragraph to give samples of $v_{\mathrm{grf}}$ and $v_{\mathrm{esc}}$. We can thus quantify the probability of a star being unbound by the fraction of samples where $v_{\mathrm{grf}}>v_{\mathrm{esc}}$.

To make this quantification rigorous we apply the \citet{brown_interval_2001} methodology. The posterior on the probability of a star being bound after $N$ trials with the star being bound in $N_{\mathrm{b}}$ trials is a $\operatorname{Beta}(N_{\mathrm{b}}+\frac{1}{2},N-N_{\mathrm{b}}+\frac{1}{2} )$ distribution. The one-sigma confidence interval centered on the median is thus easily calculable numerically.

In the subset of cases where we do not have either the proper motions or radial velocity then we assume that the missing component(s) exactly cancels the sampled solar reflex, which is equivalent to calculating the minimum Galactocentric rest-frame velocity. In this case the bound probability can be interpreted as an upper limit on the true bound probability.

One small caveat of using Gaia DR2 parallaxes is that \citet{2018arXiv180409376L} identified a global parallax offset of $-0.029 \;\mathrm{mas}$. We have accounted for this offset in our analysis. The inclusion of this offset causes stars with proper motions to become slightly more likely to be bound, however this effect is sub-dominant to the other uncertainties that we account for when calculating the bound probability.

\subsection{Automatic querying of Gaia and other catalogues}\label{sec:othercatalogues}
To ensure the catalogue incorporates the latest measurements of each star, we automatically query against large astronomical catalogues such as Gaia, SDSS and PPMXL. The querying of astrometric and photometric catalogues uses the \textsc{Astroquery} affiliated package of the \textsc{astropy} \textsc{Python} framework. The line-of-sight extinction to each star $E(B{-}V)$ is obtained from the \textsc{dustmaps} package which allows us to query the \citet{green_three-dimensional_2015,green_galactic_2018} dust maps for stars with measured distances and lying in the Pan-STARRS footprint and the \citet{schlegel_maps_1998} dust map for the other candidates. The querying of external catalogues is done assuming a cross-match radius of $\mathrm{Min}(2\;\mathrm{arcsec},3\times10\;\mathrm{yr}\times \mu_{\mathrm{tot}})$, where $\mu_{\mathrm{tot}}$ is the total proper motion of the candidate. Within this search radius we take the nearest neighbour. We also query the stars against \textsc{SIMBAD} \citep{wenger_simbad_2000} to obtain other aliases that the stars may have, which will allow users to access the catalogue independent of their preferred naming convention.

\subsection{The fast star graveyard}
It is standard practise among the Open Astronomy Catalogs to split off objects which are no longer of interest, for instance transients falsely identified as supernovae are split off from the main Open Supernova Catalog. This practise is known as putting an object in the `graveyard'. In the OFSC, this can be interpreted as a statement that a fast star is highly unlikely to be unbound and thus should not be considered to be a hypervelocity candidate. The criteria for putting a star in the graveyard is that i) each of the six kinematic components have been measured, ii) the star has 5D astrometry from Gaia DR2, and iii) the star was bound in all of the $10^4$ samples. Note that a star being in the graveyard does not mean that it has been deleted and it will be possible for a star to be resurrected as new data is obtained, for instance when Gaia DR3 is released. The only practical result of a star being in the graveyard is that it is not shown in the main section of the OFSC. 159 previously-claimed hypervelocity candidate stars were in the OFSC graveyard as of 01/06/2018.


\bsp	
\label{lastpage}
\end{document}